\documentclass[9pt,a4paper]{article}
\usepackage[a4paper,left=1.5cm,right=1.5cm,top=2cm,bottom=2cm]{geometry}
\usepackage{libertine}
\usepackage{authblk,multicol,cite,float,comment}
\usepackage{amsmath,amssymb,amsfonts}
\usepackage[
   italic,
    symbolgreek
    ]{mathastext}
\usepackage{algorithmic}
\usepackage{graphicx}
\usepackage{textcomp}
\usepackage[utf8]{inputenc}
\begin{document}

\title{Secure Architectures Implementing Trusted Coalitions for Blockchained Distributed Learning (TCLearn)}

\author[1]{Sébastien Lugan PhD}
\author[1]{Paul Desbordes PhD}
\author[2]{Luis Xavier Ramos Tormo}
\author[1]{Axel Legay PhD}
\author[1]{Beno\^it Macq PhD}

\renewcommand\Affilfont{\small}

\affil[1]{ICTEAM, Université catholique de Louvain, Louvain-la-Neuve, Belgium}
\affil[2]{Massachusetts Institute of Technology, Cambridge, MA, USA}

\date{}


%



\noindent\makebox[\textwidth][c]{
    \begin{minipage}{16cm}
\maketitle
\end{minipage}}
\noindent\makebox[\textwidth][c]{
    \begin{minipage}{15cm}
Distributed learning across a coalition of organizations allows the members of the coalition to train and share a model without sharing the data used to optimize this model. In this paper, we propose new secure architectures that guarantee preservation of data privacy, trustworthy sequence of iterative learning and equitable sharing of the learned model among each member of the coalition by using adequate encryption and blockchain mechanisms. We exemplify its deployment in the case of the distributed optimization of a deep learning convolutional neural network trained on medical images.
\footnote{This work has been partially performed during eNTERFACE'18. It has been partly funded by Wal-e-Cities, a research project co-funded by the European Regional Development Fund (ERDF) and Wallonia, Belgium.}
\end{minipage}}

~\\


\begin{multicols}{2}
{\raggedright \section{Introduction}}
\label{sec:introduction}
{D}{eep} learning algorithms, such as deep convolutional neural networks (CNN) \cite{lecun2015}, are efficient tools to help data scientists process the huge amounts of potentially very sensitive data generated e.g. in medical, financial or industrial applications (including data generated from smart sensors such as smart meters and other IoT sensors for smart cities and industry 4.0, etc.). Practitioners, for instance, routinely use such techniques for identifying various pathologies \cite{lambin2013}. This machine learning algorithm progressively extracts high level features from raw inputs thanks to multiple layers. The quality and size of the datasets used for the training phase have a major impact on performances. However, it may be difficult for a single organization such as a health center to gather enough data on its own, and multi-center studies are often hampered due to legal or ethical issues \cite{doshi2012}.

For these reasons, distributed learning \cite{boyd2010} has been suggested for multiple applications including the medical field \cite{deist2017}\cite{jochems2016}. This approach facilitates the cooperation through coalitions where each member keeps control and responsibility over its own data (including the accountability for privacy and consent of the data owners such as patients). Batches of data are processed iteratively to locally feed a shared model. Parameters generated at each step are then sent to the other organizations to be validated as an acceptable global iteration for adjusting the model parameters. Thus, partners of the coalition will jointly optimize a shared model by dividing the learning set into batches corresponding to blocks of data provided by the members of this coalition.

The naive use of a CNN in a distributed environment exposes it to a risk of corruption (intentional or not) during the training phase because of the lack of monitoring of the training increments and the difficulty to control the quality of the training datasets. The distributed learning could be monitored by a centralized certification authority which would be in charge of the validation of each iteration of the learning process. Alternatively, a blockchain could be used to store auditable records of each and every transaction on an immutable decentralized ledger. This approach has been suggested in \cite{kshetri2018}, where it is advocated that blockchains can be used to store signatures of patient records. In our context, the blockchain approach would provide distributed learning with a more robust and equitable approach for the different stakeholders involved in the learning process since all of them are involved in the certification process. Blockchain-based distributed learning has been described by Zhou et al. \cite{zhou2018}. Their architecture, called BeeKeeper 2.0, is deployed on Hyperledger Fabric and uses Hyperledger Caliper for performance testing. Weng et al. \cite{weng2018} proposed DeepChain as an algorithm based on blockchain for privacy-preserving deep learning training. It allows to perform a massive local training of intermediate gradients and to securely aggregate them among distrustful owners. DeepChain provides the auditability and confidentiality to locally train gradients for each participant while employing economic incentives to promote honest behavior. Nevertheless, it does not prevent data exposure through a malicious partner.

The challenge is to derive a new model of coalitions with a high degree of reliability, respecting data privacy and incentivizing the participation in the coalition without a central authority. In this article, we propose novel scalable security architectures, called Trusted Coalitions for distributed Learning (TCLearn), based on either public or permissioned blockchains to provide distributed deep learning with increasing levels of security and privacy preservation.

In our approach, a CNN model is shared among the members of the coalition and is optimized in an iterative sequence, each member of the coalition sequentially updating it with new batches of local data. Each iteration of the shared model is validated by a process involving the members of the coalition and is stored in the blockchain. Each step of the evolution of the model can be retrieved from the immutable ledger provided by the blockchain.

While early implementations of blockchains (such as Bitcoin) initially relied on a proof-of-work \cite{dwork1993} consensus mechanism, others have been suggested since, such as proof-of-stake \cite{king2012}. Our approach has a different goal: we aim at providing an iterative certification process for each learning step of the shared model, all of them being registered in the underlying ledger. We therefore suggest a new consensus mechanism integrating performance evaluation using block validation as part of our suggested Federated Byzantine Agreement (FBA).

In this article, the model designates a CNN architecture with its associated weights (parameters). The gradients represent the evolution of the model weights after a training step. The supervisor is an entity handling the storage of the model and controlling its access.

{\raggedright \subsection{Distributed learning for a medical application: an example}}
To illustrate the use of distributed learning, we propose to apply it on a medical challenge that has already been solved using CNN: bladder contouring on computed tomography scans \cite{leger2018}\cite{brion2019}. L\'eger et al. \cite{leger2018}\cite{brion2019} proposed to use U-Net, inspired by \cite{ronneberger2015}, to segment the bladder of 339 patients suffering from prostate cancer. Their semi-automatic approach takes two channels as input (a 3D volume of the bladder with one of its slices manually labeled by an expert) and outputs a prediction for the target tile bladder segmentation.

We reproduced their results using parameters and database used by L\'eger et al. \cite{leger2018}\cite{brion2019}. In addition, we split the initial database into several subsets to simulate several partners. Thus, the training was performed in two different ways: a centralized training using all the training samples in one run and a distributed learning successively using the smaller datasets created. Once the training over, the accuracy achieved with a distributed database is not significantly different from a centralized training (88.4\% and 88.7\%, respectively).

This simple medical use case constitutes an exemplary application of distributed learning that raises several security challenges which are listed in the next section.

{\raggedright \subsection{Security requirements in TCLearn}}
Four challenges have been identified.

\textbf{Challenge 1: protection of the model against degradation by training on inadequate data}\\
The model is exposed to a risk of corruption or degradation during the training phase. The evolution of the model must be resilient against malicious or clumsy actions to ensure increasing performances. For instance, a partner could attempt to train the model from corrupted data or from a different pathology than the studied one. The proposed approach must detect this kind of misuse and reject the resulting increment.

\textbf{Challenge 2: data privacy of the dataset used for the training}\\
Even if distributed learning allows to share a model within a consortium while keeping control over the data, it might be possible to reconstruct part of the training set from the generated gradients. This phenomenon is called long term memory effect \cite{song2017} and should be avoided.

\textbf{Challenge 3: confidentiality of the model and the gradients}\\
If the learned model has to be confidential, it is necessary to protect it and its gradients from any potential deliberate or accidental leak outside the consortium.

\textbf{Challenge 4: traceability of the model}\\
The blockchain keeps track of each and every modification to the model, but does not prevent unauthorized use of the shared model outside the coalition. Traceability of every operation involving the model must be ensured to deter any malicious event.

{\raggedright \section{A Scalable Security Architecture for Trusted Coalitions}}
Articles such as \cite{elahi2009} give an illustration of the well-known trade-off issue between security and cost. In this section, we will develop three different methods corresponding to three distinct trust levels depending on the shared rules in the coalition:
\begin{itemize}
\item \textbf{Method TCLearn-A:} the learned model is public but each member of the coalition is accountable for the privacy protection of his own data.
\item \textbf{Method TCLearn-B:} the learned model is private (shared only within the coalition) and the members of the coalition are trusting each other.
\item \textbf{Method TCLearn-C:} the members of the coalition do not trust each other and want to prevent any unfair behavior from any of them, such as unauthorized use or leak of the model outside the coalition.
\end{itemize}

These three methods address the previously described security issues at different levels (see Table 1), offering an inherent tradeoff between security needs and costs.
\end{multicols}

\begin{table}[thpb]
    \centering
    \begin{tabular}{c|c|c|c|c}
          & Data & Protection against & Model & Model  \\
          & Privacy & degradation & Privacy & traceability  \\
          \hline
          TCLearn-A & ++ & ++ & - & - \\
          TCLearn-B & ++ & ++ & + & + \\
          TCLearn-C & ++ & ++ & ++ & ++ \\
          \hline
    \end{tabular}
    \caption{Summary of the features for the three TCLearn methods \newline (-: absent, +: basic, ++: developed)}
    \label{tab:summary}
\end{table}

\begin{multicols}{2}
{\raggedright \subsection{Architecture TCLearn-A}}
\noindent
\textit{Type of coalition:} coalition sharing a public model built using private datasets. The integrity of the increments is ensured through the use of a new FBA protocol.

\textbf{Mechanism to guarantee challenge 1} (protection of the model against degradation by training on inadequate data)\\
Our approach relies on a blockchain to carry unalterable cryptographic hashes of the successive training steps of a model built in a distributed environment. The iteratively optimized model is made public. Each block represents an iteration step achieved locally by a specific member of the coalition and validated by the whole coalition. First, the model and the genesis block are initialized, setting the architecture (layers, activation functions, loss function, etc.) and the weights according to a normal distribution. The weights of the model are updated iteratively by the batches of data provided by the members of the coalition.

In our approach, we use a blockchain relying on a federated byzantine agreement to prevent corrupted increments caused by inadequate training to be added to the model. The candidate increment has to be validated by multiple validators. The role of the FBA is to validate the concatenation of a new block to the chain. Since the blockchain and the deep learning model are strongly linked in TCLearn, the FBA has two goals:
\begin{itemize}
    \item Control the quality of the updated CNN model;
    \item Control the integrity of the new block (hash, index, timestamp, etc.).
\end{itemize}

The FBA process starts with the random selection of validators within the consortium. The chance to be selected depends on the size of the consortium and the proportion of data brought by each member (Equation \ref{eq_stren}).
\begin{equation}
    S_i = 1/N + D_i/D
\label{eq_stren}
\end{equation}

where $S_i$ is the strength of the partner $i$, $N$ the total number of partners in the consortium, $D$ the total amount of samples used for the model and $D_i$ the amount of samples supplied by the partner $i$. Initially, this strength is the same for each of the partners and evolves with each contribution. The main role of the validators is to check the candidate model $M_{i+1}$ incremented from $M_i$ proposed by a partner. $M_{i+1}$ must show an improvement over the previous model. Enforcing this improvement prevents a degradation of performances caused by the presence of corrupted data in the training set.

Two types of test databases are used to assess the performances of each increment of the model (Fig. 1) and to avoid the introduction of invalid or inadequate training sets:
\begin{itemize}
    \item A global test database ($G$), common for every block creation and for all the partners. This database is created by experts to be representative of the pathology.
    \item A local test database ($L$), different for each partner in the consortium. To avoid overfitting on the global test dataset ($G$), a small percentage of the input signals is locally put aside for each contribution. It is later used by each validator as a local test set to individually evaluate the proposed model.
\end{itemize}

Both datasets are used for the testing phase. First, results obtained using the common, global dataset are compared between the validators to ensure that the candidate model is functional and identical. Then, those ``global'' results are merged with those obtained using the individual, local datasets. To be accepted, the candidate model must have higher performances than the previous model within a specific threshold $\lambda$ ($\in$ [0, 1]) (Equation \ref{eq_perf}).
\begin{equation}
    \textrm{Block creation IF } \lambda * perf(M_i) > perf(M_{i+1})
\label{eq_perf}
\end{equation}

\begin{figure}[H]
    \includegraphics[width = 3in]{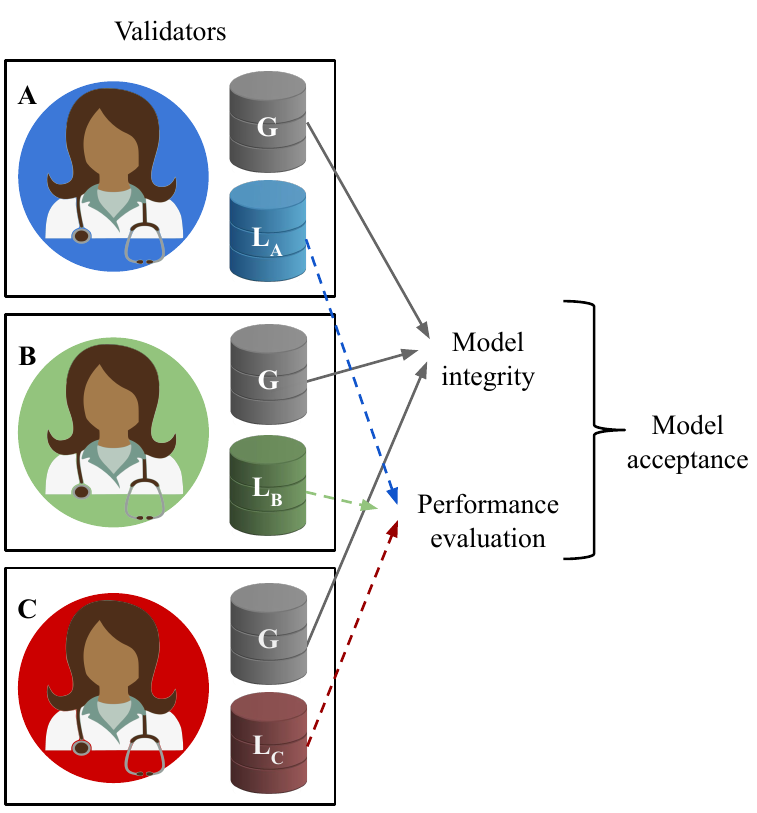}
    \caption{Federated byzantine agreement and candidate model checking process. Two datasets are used: a global one ($G$), similar for all the partners, used for the control of the model integrity and a local one ($L$), different for all the partners, used for the performance evaluation. After a majority vote, the candidate model is accepted or not.}
\label{fig:val}
\end{figure}

Once the model is accepted, a new block can be created. First, the block creation requires to randomly select a ``general'' (or ``speaker'') among the validators. The ``general'' will be the creator of a new block containing the reference to the validated model $M_{i+1}$ and the ID of the partner who proposed it. Every validator then checks the integrity of this candidate block.

This block is analyzed in the frame of a delegated byzantine fault tolerance system. Each validator broadcasts its opinion (acceptance or rejection) regarding this block to the other validators. If at least two thirds of the validators agree, the FBA process can stop, leading to the acceptance or the rejection of the block. If not, the role of ``general'' is switched to another randomly selected validator and the block creation process can restart. If the block is accepted by the validators, the ``general'' can append it to the blockchain and broadcasts this update to the whole consortium, requesting synchronization of the blockchains.

The global scheme of TCLearn-A is represented Fig. 2.

\begin{figure}[H]
\centering
    \includegraphics[width = 3in]{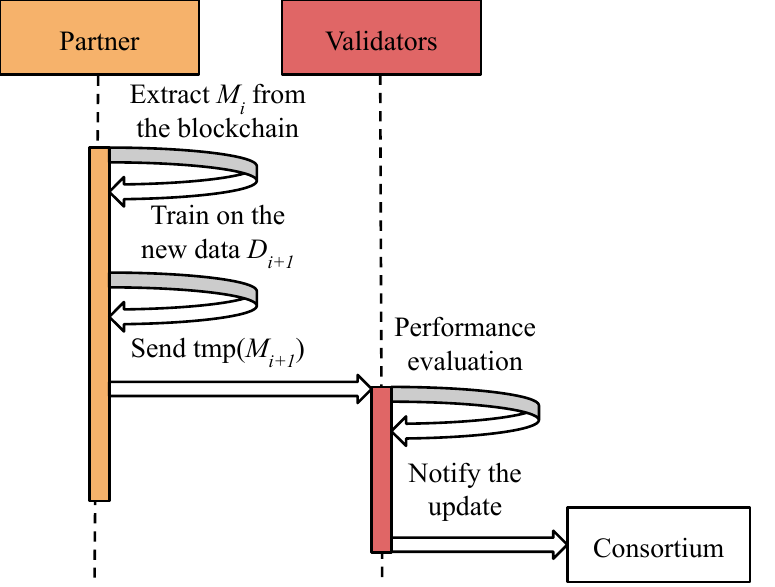}
    \caption{Scheme of the TCLearn-A procedure. Partner trains a model ($M_i$) on their data ($D_{i+1}$). The candidate model $tmp(M_{i+1})$ is validated by federal byzantine agreement.}
\label{fig:methA}
\end{figure}

Each block includes:
\begin{itemize}
    \item Block index
    \item Timestamp
    \item Previous block hash
    \item Hash of the model's parameters
    \item User ID: identification of the contributor
    \item Users' strength: contribution level of every member for the FBA
    \item Block hash
\end{itemize}

\textbf{Mechanism to guarantee challenge 2} (data privacy of the dataset used for the training)\\
A partner willing to improve a model first fetches the current version of the model, trains it locally with its own datasets and finally uploads the resulting gradients. This way, the datasets used for training never leave the partner's infrastructure, ensuring their privacy and excluding any leakage of the processed data. After this training, the generated gradients $G_{i+1}$ are uploaded and merged with the previous model $M_i$, leading to a candidate model, noted $tmp(M_{i+1})$.

Several approaches have been proposed in the literature to mitigate the long term memory effect. As an example, adding some noise to the uploaded gradients has been considered in \cite{shokri2015}. This leads to a trade-off between data privacy and training performances. Other approaches rely on Homomorphic Encryption (HE) techniques \cite{phong2018}. Abadi et al. \cite{abadi2016} established a link between the size of the dataset divided in batches and the confidentiality on the original data. Consequently, we propose to iterate the learning model by batches with a minimum size.

{\raggedright \subsection{Architecture TCLearn-B}}
\noindent
\textit{Type of coalition:} coalition sharing a private model built using private datasets among a restricted consortium of trusted partners. The integrity of the increments is ensured through the use of FBA protocol. In this situation, the model has to be protected during transfers between partners and for its storage.

\textbf{Mechanism to guarantee challenges 1 \& 2}\\
This approach being based on TCLearn-A, the data privacy of the inputs is preserved and the model iterations are validated by members of the coalition.

\textbf{Mechanism to guarantee challenge 3} (confidentiality of the model and the gradients)\\
With TCLearn-A, the evolutions of the model are certified by the blockchain. However, this scheme does not offer confidentiality guarantee on the model during its distribution to partners. This situation is not acceptable in some situations where the privacy of the model should be preserved (for example if the members of the coalition would like to avoid any leakage of the model).

To solve this issue, the storage, transfer and upgrade through gradient computations of the model have to be protected by encryption, where the private keys are stored by some trusted entities. In this work, we propose to isolate all iterations of the model in an external, off-chain encrypted storage (``vault'') and control its access by each partner.

We also suggest the use of secure transport (e.g. TLS or S/MIME, cf. section 3 for details) for transferring the model. Moreover, the model could be securely stored using an efficient encryption method and by implementing access control and auditing mechanisms (see section 3 for details). Only authorized users should be able to download a given version of the model weights, and each access should be logged into an audit trail.

\begin{figure}[H]
\centering
    \includegraphics[width = 3in]{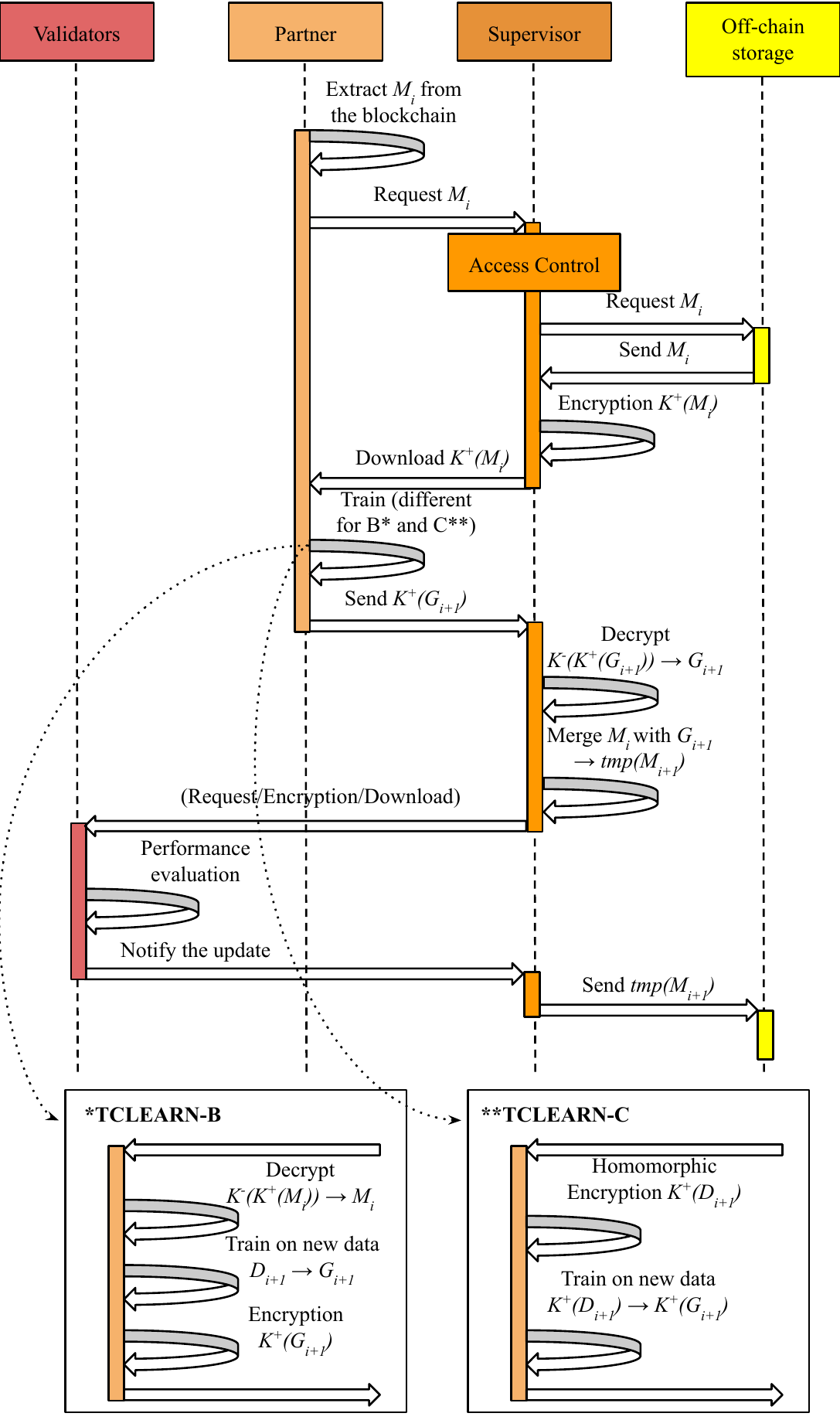}
    \caption{Scheme of the TCLearn-B \& C procedure. The partner trains a model ($M_i$) on their data ($D_{i+1}$) leading to new gradients ($G_{i+1}$). The model is successively encrypted and decrypted, e.g. using a public ($K^+$) and a private key ($K^-$), respectively. The candidate model ($tmp(M_{i+1})$) is validated by federated byzantine agreement. The two methods are principally distinct by the training process in or out of the encryption domain.}
\label{fig:methBC}
\end{figure}

\textbf{Mechanism to guarantee challenge 4} (traceability of the model)\\
In order to (1) ensure that only authorized participants are granted access to the models and (2) to minimize the risks of leak, we propose to store the actual models (including the associated weights) in an external, secure ``vault'' (off-chain storage).

In this off-chain storage approach, the blockchain only provides ``links'' to the corresponding version of the model. This introduces a single point of failure (the secure, off-chain storage and its associated access control infrastructure), but offers three major advantages. First, it greatly reduces the size of the blockchain while increasing its scalability: each participant needs to synchronize fewer data. Second, this approach permits to implement an active access control to all of the stored information (e.g. who is allowed to access which kind of data, when and how). Finally, each of the requests to access the secure, off-chain storage could be recorded in a journal, offering the ability to audit all accesses to any record. This audit could possibly be used to gather statistics about the actual accesses of each individual user to models. This in turn could be used to restrain future accesses (e.g. allowing user to access a certain number of models depending on their level of contribution to the models). Therefore, a moderated level of traceability is offered: in case of leak of a given version of a model, it is possible to audit all of the requests related to this specific version predating the leak in order to establish a list of partners which actually downloaded this version and potentially leaked it.

The blockchain/offline storage approach requires an entity (the ``supervisor'') managing the access control and the secure storage of the models (see section 3 for details). An overview of this approach is presented in Fig. 3.

{\raggedright \subsection{Architecture TCLearn-C}}
\noindent
\textit{Type of coalition:} coalition sharing a private model built using private datasets among a restricted consortium of untrusting partners. The integrity of the increments is ensured through the use of FBA protocol. To protect the model in this situation, it is necessary to secure the exchanges and the storage even at the partners' facilities.

\textbf{Mechanism to guarantee challenges 1 \& 2}\\
This approach being based on TCLearn-A, the data privacy of the inputs is preserved and the model iterations are validated by members of the coalition.

\textbf{Mechanism to guarantee challenges 3 \& 4}\\
In this scenario, our objective is to identify the partner responsible for the leakage of a model. Such identification could be performed by adding some unique, hidden information to the model provided to each partner, e.g. by altering the weights by introducing a moderate noise following a specific, hidden pattern (constituting a watermark) every time the model is requested by a partner. Both the date of the access, the identity of the partner and the associated hidden pattern (the watermark) could then be stored in an audit trail allowing to uniquely identify the partner associated with a leaked model. Unfortunately, CNN are quite robust to a slight alteration of the weights, which means that an adverse party might try to alter those (watermarked) weights, jeopardizing the recovery of the watermark while keeping the model usable. This adverse party could even perform a subsequent training of the model on new datasets, further compromising the recoverability of the watermark.

Another option could be to send to each partner a model encrypted with a specific key, allowing them to manipulate (use and even train) it while being unable to decrypt it. Some crypto algorithms (such as the ElGamal scheme \cite{elgamal1985}) allow operations to be performed directly on the encrypted data without knowing the associated private key. For instance, the ElGamal scheme offers an encrypted product operator allowing to compute the product $E(m_1 \cdot m_2) = HM(E(m_1), E(m_2))$ of two encrypted messages $E(m_1)$ and $E(m_2)$. An operator offering such a property constitutes a homomorphic operator. An encryption algorithm offering both homomorphic product and homomorphic addition, such as the Brakerski-Gentry-Vaikuntanathan scheme \cite{brakerski2012}, constitutes a fully homomorphic encryption (FHE) scheme.

Homomorphic encryption brings a new level of protection to the model (since actually only the supervisor is able to decrypt it), but also a considerable drawback: in order to perform prediction, the encrypted result must be send by the partner to the supervisor to be decrypted and the final result sent back to the partner. This reduces the autonomy of the partners (depending on the supervisor to perform any prediction using the model) and potentially introduces a single point of failure.

Similarly to TCLearn-B, the confidentiality and the distribution control over the models is ensured by off-chain storage of the models and gradients. However, since the members of the consortium do not trust each other, no member should have access to unencrypted models or gradients. This is ensured by the use of the aforementioned homomorphic encryption technique. In order to use the model (prediction), the input signals must first be encrypted using the same technique and the same homomorphic public key (which must be sent together with the encrypted models and gradients to each partner). This public key could only be used to encrypt (and not decrypt) data. Once the prediction operation (using encrypted model and input signals) performed, the result must be decrypted, which must be performed by the supervisor (which holds both the homomorphic public and private keys corresponding to the models sent to each partner). The supervisor uses its homomorphic private key to decrypt the result, which is then sent to the requesting partner. The supervisor must carefully inspect the ``results'' to be decrypted in order to reject any attempt to decrypt models or gradients.

{\raggedright \section{TCLearn implementations considerations}}
{\raggedright \subsection{Additional features}}
In addition to the TCLearn approaches, we propose two optional extra features.

\textbf{Reverting to a previous state of the model}\\
Each and every increment of the model recorded on the blockchain and validated using the FBA are accessible by the partners. In case of a late detection of corruption, it could be necessary to restore the model to one of its previous states. For this reason, we add the possibility to perform prediction or even continue the training from old weights stored in the off-chain storage. Only authorized partners are able to perform this operation, in which case a new block is generated, outshining previous learning steps.

\textbf{Updates to the model hyperparameters}\\
The model architecture (kernel size, number of layers etc.) is initialized in the genesis block of the blockchain but some hyperparameters can be modified during the learning step (epoch number, batch size). The learning rate of the optimizer for the gradient descent is a critical parameter. It could be high at the beginning of the blockchain because the model is still naive, but the more precise, the lower the learning rate should be (without becoming too low). A method that automatically adapts the learning rate according to the performances of the model could be integrated into our concept \cite{duchi2011}.

\textbf{Audit \& traceability of the training data}\\
The evaluation of performances may not be enough to avoid duplicate input signals during training. For this reason, anonymous IDs can be attached to each input signal and stored in the off-chain storage. This process also enforces the auditability and the traceability of the input signals. This can be implemented as an additional field stored in each block, such as an anonymized data ID, therefore avoiding training on duplicate input signals.

{\raggedright \subsection{Secure authentication and transport of the model}}
The secure authentication and transport of the model could be performed in two ways: either online (requiring direct, interactive communication between the partner and the supervisor) or offline (allowing for delayed communication, e.g. using periodic batches of file transfers or using e-mail).

In the former case, we strongly recommend using an industry-standard protocol such as TLS (Transport Layer Security) v.1.3 (RFC 8446), which is used for instance in HTTPS. In contrast with network-level encryption (such as IPsec VPNs), TLS offers end-to-end encryption that guarantees the confidentiality of the model and gradients, including between the secure gateway (e.g. VPN concentrator) and the machine performing the machine learning. The server (the supervisor) must be authenticated by validating its X.509 certification chain. The identity of the client could also optionally be requested (using a X.509 certificate chain) in order to provide a stronger authentication mechanism. Once the TLS handshake and authentication successfully performed, the model and its gradients could be safely exchanged using strong encryption (such as AES-256).

In the latter case, we suggest to use industry-standard protocol such as S/MIME (RFC 5751). Both the supervisor and the partner need to possess X.509 certificates, which could be used to both authenticate (digital signature of the request from the partner and of the reply from the supervisor) and encrypt the data (using the public key of the partner so that only this partner could decrypt it). In this scenario, the partner could send a request, digitally signing it using its X.509 private key; the supervisor ensures the authenticity of the request by verifying the provided digital signature (decrypts it using the partner's public key). Once the partner (and the correspond request) authenticated, the supervisor could send the encrypted model and gradients. This operation could e.g. be performed by the supervisor by first generating a random session key which could be used as a symmetric key with a symmetric encryption system such as AES. This random session key is then encrypted using the partner's public key (so that only him/her could decrypt it using his/her private key) and sent along with the encrypted model and gradients to the partner. The partner then simply needs to decrypt the symmetric session key using his/her private key, and then using this (decrypted) session key in order to decrypt the message. This scheme (illustrated in Fig. 3) requires each party to possess an X.509 certificate and the associated private key in order to decrypt information provided by other parties.

{\raggedright \subsection{Considerations regarding the use of homomorphism in CNN}}
In TCLearn-C we suggest the use of homomorphic encryption to ensure traceability and confidentiality of the model (cf. section 2.C). In this scenario, the whole manipulation of the CNN takes place in the homomorphic domain, which introduces some challenges that are described hereafter.

First, the neural network algorithm requires to use and combine both addition and multiplication operations. Consequently, the ``somewhat homomorphic encryption'' scheme (see ElGamal \cite{elgamal1985}, BGV \cite{brakerski2012} or Paillier \cite{paillier1999}), which allows the use of one sole type of homomorphic arithmetic operation, cannot be used. We thus have to use FHE, which requires considerable memory and computational resources.

The use of CNN in the homomorphic domain also introduces implementation issues. Indeed, while most FHE implementations only support homomorphic addition and homomorphic multiplication, implementation of CNN activation functions in the homomorphic domain requires complex operations such as trigonometric (tanH), exponentials (sigmoid), or tests (ReLU). Zhang et al. \cite{zhang2016} proposed the use of a Taylor development to replace the sigmoid function in order to compensate the lack of exponential function in FHE schemes. Albeit this method allows us to use the sigmoid function in the homomorphic domain, it remains approximate. Moreover, it still considerably increases the amount of operations to be performed.

Solving those problems has been the subject of some recent work. As an example, in \cite{bourse2018}, Bourse et al. presented a framework for homomorphic evaluation of neural networks using a highly optimized FHE algorithm. This scheme, named TFHE, offers several orders of magnitude performance improvements over previous FHE architectures. In this article, the authors used of a ``discretized'' neural network to allow the creation of a model capable of fitting data from the MNIST database. Several tips are proposed to reduce the time required for learning and prediction (bootstrapping, look-up table, noise management).

One could thus use Bourse's solution. If this approach is not sufficient to mitigate the resource issues associated with TCLearn-C, the training process could alternatively be performed using tamper-proof black boxes deployed in each of the partner's facilities. Such black boxes could be used to decrypt the CNN model, retrain it, and then re-encrypt it without anyone being able to interfere. However, this solution requires all of the partners to request the installation and maintenance of this black box by a trusted third party service provider.

{\raggedright \section{Conclusions}}
In this article, we propose a new architecture for distributed learning of a model based on a mechanism of federated byzantine agreement. The performance of the model is ensured through a shared evaluation of individual contributions leading to the acceptance or rejection based on an objective criterion.

 This approach allows to constitute trusted coalitions in which the actions for updating the model by the members are registered on a public ledger implemented as a blockchain. We have explored three kinds of coalitions depending on the access control required for the distribution of the model. Each approach corresponds to distinct trust levels depending on the shared rules in the coalition. We have proposed solutions that rely on efficient cryptographic tools including homomorphic encryption. We have exemplified our proposed architectures for the case of the distributed learning of a CNN model applied to distributed medical images databases. The proposed architectures allow keeping data privacy thanks to a system of encryption and off-chain storage avoiding the dissemination of sensitive medical data or metadata.

\end{multicols}

\end{document}